\pdfoutput=1
\documentclass[a4paper,12pt]{article}
\usepackage{latexsym}
\usepackage{amssymb}
\usepackage{amsmath}
\usepackage{mathrsfs}
\usepackage{graphicx}
\usepackage{tikz}
\usepackage{hyperref}

\newcommand{\be}{\begin{equation}}
\newcommand{\ee}{\end{equation}}
\numberwithin{equation}{section}

\begin{document}

\begin{titlepage}

\begin{center}
{\Large {\bf Induced Boundary Flow on the $c=1$ Orbifold Moduli Space.}}\\\vspace{1in} {\large S. Elitzur \footnote{elitzur@vms.huji.ac.il}, B. Karni \footnote{boaz.karni@mail.huji.ac.il}, E. Rabinovici \footnote{eliezer@vms.huji.ac.il}}\\ \vspace{0.5in}
{{\it Racah Institute of Physics, The Hebrew University, \\ Jerusalem 91904, Israel}}\\
\end{center}

\begin{abstract}
Boundary flow in the $c=1$ $2d$ CFT of a $\mathbb{Z}_2$ orbifold of a free boson on a circle is considered. Adding a bulk marginal operator to the $c=1$ orbifold branch induces a boundary flow.
We show that this flow is consistent for any bulk marginal operator and known initial given boundary condition.
The supersymmetric $c=\frac{3}{2}$ case is also mentioned.

For the circle branch of the moduli space this has been shown in arxiv: 0609034v2 [hep-th].

The ground state multiplicity ($g_b$) is calculated and it is shown that it does indeed decrease.\end{abstract}
\end{titlepage}

\tableofcontents

\section{Introduction}
String theory presents us with a multitude of vacua and with no criteria to choose among them. Maybe one is destined to remain with these vacua. Some of them, though, contain tachyons and are unstable. The search for possible stable vacua resulting from the flow induced by closed string tachyons is still not over {\cite{[1]}. In world sheet language this instability is connected to the RG flow induced by bulk relevant operators in the internal CFT corresponding to a given background. In presence of branes there are also boundary operators. Turning on a boundary relevant operator induces a flow of the boundary states in the system. Such flows  can be controlled and do lead to stable vacua \cite{[2],[3]}.

 In \cite{[4]} a different kind of flow mechanism was pointed out, one in which turning on a marginal operator in the bulk theory induces a flow in the boundary theory from one boundary state to another. This flow is necessary because for a free boson on a circle, or an orbifold, that has a compactification radius that is equal to the self-dual radius or a rational multiple thereof, the boundary moduli space is much larger than for a compactification radius that is an irrational multiple of the self-dual radius. Consequently if the initial state of the system is a boundary state that does not exist for an irrational compactification radius, any change of the radius would necessarily result in a flow to an allowed boundary state. For this flow to be consistent with the boundary moduli space structure, various constraints should be obeyed. In \cite{[4]} some of these constraints were indeed tested for that part of the $c=1$ of the bulk moduli space corresponding to compactification circles of different radii.  Orbifolds play an important role in the string landscape. In this paper those flows induced on the $c=1$ orbi-circle branch are thus studied ( \cite{[5]}, \cite{[6]} and \cite{[7]} ).\\
Several features of this branch which were discussed in \cite{[8]} are used and the induced flows on the boundary are mapped. The detailed way in which consistency is maintained in all possible induced flows on the boundary is studied. We also check that the ground state multiplicity $(g_b)$ indeed decreases along the flow. \cite{[9]}\\
The structure of the paper is as follows: In section ~\ref{sec:bosonic theory} we deal with the bosonic theory and say a few words regarding the $\mathcal{N}=1$ superconformal case. In section ~\ref{sec:boundary entropy} we evaluate the ground state multiplicity, and show that it decreases, as expected.
\section{Bosonic theory} \label{sec:bosonic theory}
In this section the induced flows of the $c=1$ case compactified on a circle shown in \cite{[6]} is reviewed. The results are extended to the $\mathbb{Z}_2$ orbifold both at, and away from, the multicritical point, which is the point where the theories of a circle and of an orbifold intersect. This happens when the circle compactification radius is  twice the self dual radius, and the radius of the orbifold is the self dual radius.

\subsection{Review}
Consider a CFT with a boundary. The beta-functions for the bulk and boundary coupling constants in a general BCFT have the form of \begin{equation}{\dot{\lambda }_{k}}=\left(2-{h_{\phi }}_{k} \right){{\lambda }_{k}}+\pi {{C}_{ijk}}{{\lambda }_{i}}{{\lambda }_{j}}+\mathsf{\mathcal{O}}\left( {{\lambda }^{3}} \right)\label{eq1}\end{equation}

\begin{equation}{\dot{\mu }_{k}}=\left( 1-{{{h}_{\psi }}_{k}} \right){{\mu }_{k}}+\frac{1}{2}{{B}_{ki}}{{\lambda }_{i}}+{{D}_{ijk}}{{\mu }_{i}}{{\mu }_{j}}+\mathsf{\mathcal{O}}\left( \mu \lambda ,{{\mu }^{3}},{{\lambda }^{2}} \right)\label{eq2}\end{equation}
Where $\phi_k$ is the bulk field, $h_{\phi_k}$ is its conformal dimension, $\lambda_k$ is its coupling constant, and $\psi_k$, $h_{\psi_k}$ and $\mu_k$ are the corresponding quantities for the boundary. The constants $C_{ijk}$ are related to the OPE of two bulk fields, $D_{ijk}$ to that of two boundary fields and $B_{kj}$ to the OPE of a bulk field with a boundary field. As was pointed out in \cite{[4]}, it follows from second term of (\ref{eq2}) that in general even if initially all boundary operators are turned off, i.e. $\mu_k=0$ for all $k$, the boundary beta-function does not vanish if the $B_{ki}$'s are not all zero. Hence when a bulk-boundary coupling exists a non-vanishing boundary coupling constant will appear, even if initially it is zero. \\

Consider first the case of $c=1$ BCFT on a circle at the self dual radius, $r_{s.d}=\frac{1}{\sqrt2}$ (we take $\alpha'=\frac{1}{\sqrt{2}}$). This theory is equivalent to a level $1$ $SU\left(2\right)$ WZW theory, (see for example \cite{[12]}). The moduli space of boundary states where the value of the radius of compactification is either that of the self-dual radius or any rational multiple of it is rather rich, being parameterized by the three dimensional $SU(2)$ group manifold. In particular it is larger than that of the more generic case when the compactification radius is an irrational multiple of the self-dual radius. In the latter case there are only two possible boundary conditions: Dirichlet or Neumann, each of which being  parameterized by a single circle. This imposes a severe constraint on any flow induced by changing the radius from a rational multiple of the self-dual radius to an irrational multiple of it, a situation which occurs with any continuous flow starting at a rational multiple radius. It was shown in \cite{[4]} that when the radius-changing marginal operator, $\Phi_{circ}=J^3\bar{J}^3$, is "turned on" with a coupling constant $\lambda$, general boundary conditions that exist only at the self dual radius or a rational multiple thereof flow to pure Dirichlet or Neumann conditions. Those exist at any radius. To show this one needs to consider the perturbation of the boundary current, $J^\gamma$, by the bulk radius changing operator, $\Phi_{circ}$, for the theory at $r=r_{s.d}$. As discussed above, the condition for the stability of the boundary under radius changing is the vanishing of the bulk-boundary coupling constant, $B_{\gamma\Phi}$, between the radius changing operator $\Phi_{circ}(z,\bar{z})$ and any marginal boundary operator $J^\gamma (x)$. To first order this is given by the 2-point function between these operators as: {\begin{equation}\left<J^\gamma\left(x\right)\left(J^3\bar{J}^3\right)\left(z,\bar{z}\right)\right>_{g}\sim B_{\gamma\Phi}|z-\bar{z}|^{-1} |x-z|^{-2} \label{eq3}\end{equation} It is calculated by replacing the boundary condition by a boundary state, $\left|\left|g\right>\right>_{r_{s.d}}$, labeled by
\begin{equation} g=\left( \begin{matrix}
   a & b  \\
   {-{b^{*}}} & {{a}^{*}}  \end{matrix}
 \right)\in SU\left( 2 \right)\nonumber\end{equation} with $|a|^2 + |b|^2=1$
such that \begin{equation}\left(gJ^{3}_mg^{-1}+\bar{J}^{3}_{-m}\right)\left|\left|g\right>\right>_{r_{s.d}}=0\label{eq4}\end{equation} where  $J^{i}_{m}$ are the Laurant modes of the currents of the WZW model. The detailed calculation appears in \cite{[4]}. One obtains that  $B_{\gamma\Phi}$ is proportional to $Tr(t^\gamma[t^3,gt^3g^{-1}])$, $t^{i}, i=1,2,3$ are defined by $t^{i}=\frac{1}{\sqrt{2}}\sigma^{i}$ with $\sigma^{i}$ the standard Pauli matrices. This means that the $J^\gamma$ whose flow is induced by $\Phi_{circ}$ corresponds to the Lie algebra element $t^\gamma\sim [t^3,gt^3g^{-1}]$. For this current the bulk-boundary coupling is proportional to
\begin{equation} Tr\left(\left[t^3,gt^3g^{-1}\right]^2\right)=-8\left|a\right|^2\left|b\right|^2\label{eq5}\end{equation}
So unless $b=0$ or $a=0$ (which correspond, respectively to pure Dirichlet or pure Neumann boundary conditions), $\dot{\mu_\gamma}$ does not vanish. Taking into account the appropriate normalization,
the resulting infinitesimal perturbation of the boundary state (\ref{eq4}) is \begin{equation}\delta g\sim -\lambda t^\gamma g=-\frac{\lambda}{\sqrt{2}}\left( \begin{matrix}  -a\left|\frac{b}{a}\right| & b\left|\frac{a}{b}\right| \\ \\  -b^*\left|\frac{a}{b}\right| & -a^*\left|\frac{b}{a}\right| \end{matrix} \right).\label{eq6}\end{equation}  Namely, $\delta a =\frac{\lambda}{\sqrt{2}} a|\frac{b}{a}|$ and $\delta b=-\frac{\lambda}{\sqrt{2}} b|\frac{a}{b}|$.  So the magnitudes of $a$ and $b$ change keeping their phases fixed. If the radius is increased (positive $\lambda$) then, to first order, $|b|$ decreases and  the state eventually flows to a Dirichlet brane \cite{[4]}, while if  the radius is decreased (negative $\lambda$) then, again to first order, $|a|$ decreases and the flow is towards a Neumann brane.

 The case of a radius $r=\frac{1}{N}r_{s.d}$ can be thought of as an orbifold of the  circle theory with $r_{s.d}$ by the identification  $X\mapsto X+\frac{2\pi r_{s.d}}{N}$ . In terms of $SU(2)$ variables the identification is $g\mapsto e^{i\frac{\pi}{N}\sigma_{3}}ge^{-i\frac{\pi}{N}\sigma_{3}}$ namely $a \mapsto a, b \mapsto e^{-i\frac{2\pi}{N}}b$. The generic boundary states are symmetric projections of the boundary states of the original $r_{s.d}$ circle, namely
 \begin{equation}
 \left|\left|g(a,b)\right>\right>_{r_{s.d}/N}=\frac{1}{\sqrt{N}}\sum_{m=0}^{N-1} \left|\left|g(a,be^{\frac{2\pi i m}{N}})\right>\right>_{r_{s.d}}. \label{eq7}\end{equation}
 The response of such a brane to a change of the radius is again encoded in the bulk-boundary coupling which, like in (\ref{eq3}), will now be the coefficient of $|z-\bar{z}|^{-1} |x-z|^{-2}$ in the 2 point function
 \begin{equation}\frac{1}{N}\sum_{m=0}^{N-1}\left<J^\gamma_{(m)}\left(x\right)
 \left(J^3\bar{J}^3\right)\left(z\right)\right>_{g_{(m)}}\label{eq8}\end{equation}
 where $J^\gamma_{(m)}$ corresponds to the generator $e^{i\frac{\pi m}{N}\sigma_{3}}t^\gamma e^{-i\frac{\pi m}{N}\sigma_{3}}$ and $g_{(m)}=e^{i\frac{\pi m}{N}\sigma_{3}}g e^{-i\frac{\pi m}{N}\sigma_{3}}$.
 According to the discussion following (\ref{eq4}) the $m$'th term in the sum is proportional to
 $Tr(e^{i\frac{\pi m}{N}\sigma_{3}}t^\gamma e^{-i\frac{\pi m}{N}\sigma_{3}}[t^3,g_{(m)}t^3g_{(m)}^{-1}])$ which is independent of $m$ and equals $Tr(t^\gamma[t^3,gt^3g^{-1}])$. We get then the same result for the bulk boundary coupling as in the case of the self dual circle. Hence we reach the same conclusion: under radius changing the brane runs either to a  brane at $b=0$ or to one at $a=0$.

 The brane at $a=0$ is a Neumann brane in the circle language. Those at $b=0, |a|=1$, which is the fixed circle of the orbifold identification, actually represent a stack of $N$ Dirichlet branes on top of each other, being the continuous limit of (\ref{eq7}) where $N$ images coincide on the group manifold.
\cite{[10]}. Indeed in this case all the vectors in the sum of (\ref{eq7}) are parallel and the norm of the boundary state is $N$ rather than $1$. This will be useful when calculating the ground state multiplicity below. The position of this stack on the circle is parameterized by the phase of $a$.

 In addition to the generic branes there are additional twisted branes, located at the fixed circle. These, in the circle language, are ordinary Dirichlet branes with multiplicity less than $N$. They can sit anywhere on the circle since in the circle language the transformation $X\mapsto X+\frac{2\pi r_{s.d}}{N}$ has no fixed point. Since these are Dirichlet branes that exist for any radius, they are stable under radius changing.

The case of  a circle with a radius $r=Mr_{s.d}$ with $M$ integer can be discussed with identical terms using $T$ duality. This duality changes the sign of the right handed momentum, keeping the left handed momentum unchanged. In terms of $SU(2)$ variables it takes the matrix $g$ to the matrix $g'=gi\sigma^{1}$  so that multiplying $g$ from the right by $t^3$ induces a multiplication of $g'$ from the right by $-t^3$. So under $T$ duality $a'=ib^*, b'=-ia^*$. For $r=Mr_{s.d}$ the dual radius gets multiplied by $\frac{1}{M}$. It is an orbifold of the $T$ dual theory by
$X'\mapsto X'+\frac{2\pi r_{s.d}}{M}$. In $SU(2)$ language it identifies $g'$ with $e^{i\frac{\pi}{M}\sigma_{3}}g'e^{-i\frac{\pi}{M}\sigma_{3}}$ which means  $a'\mapsto a'$ and $ b' \mapsto e^{-i\frac{2\pi}{M}}b'$. In terms of the original parameters $g\mapsto e^{i\frac{\pi}{M}\sigma_{3}}ge^{i\frac{\pi}{M}\sigma_{3}}$ namely  $a\mapsto e^{i\frac{2\pi}{M}} a$ and $b\mapsto b$. By the same arguments as before generic branes, which are symmetric projections of those of the self dual circle, will flow under radius changing either towards  Dirichlet branes at $b=0$, or  towards stacks of $M$ Neumann branes at the fixed circle $a=0,|b|=1$.  The twisted states are individual Neumann branes sitting on the fixed circle $a=0, |b|=1$ so they are stable under radius changing.

The case of a general rational circle with $r=\frac{M}{N}r_{s.d}$ is an orbifold of the self dual circle under both transformations, a shift by $\frac{2\pi r_{s.d}}{N}$ and a dual shift by $\frac{2\pi r_{s.d}}{M}$. The same reasoning shows that generic branes flow under radius changing either to a stack of $N$ Dirichlet branes or to a stack of $M$ Neumann branes. The two transformations have no common fixed point. The twisted branes are either ordinary dirichlet branes or ordinary Neumann branes which exist for every radius.

\subsection{The $\mathbb{Z}_2$ orbifold}
Consider first the $\mathbb{Z}_2$ orbifold at the self dual radius. The orbifolding operation identifies fields on opposing points of the circle: $X\left(z,\bar{z}\right)\mapsto-X\left(z,\bar{z}\right)$.
  Before considering  the large moduli-space at the self-dual orbifold, recall first the branes available at any radius \cite{[8]}. These are orbifold-invariant combination of the pure Dirichlet and pure Neumann states of the circle. For  Dirichlet these would be \begin{equation}\left|\left|D;\theta\right>\right>_{orb}=\frac{1}{\sqrt{2}}\left(\left|\left|D;\theta\right>\right>_{circ}+\left|\left|D;-\theta\right>
  \right>_{circ}\right)\label{eq9}\end{equation} At the two fixed point of the orbifold $\left(\theta_0=0,\pi\right)$ there is, in addition to these, a twisted sector:
\begin{equation} \left|\left|D;\theta_0\right>\right>_{tw}=e^{\sum_{r=1/2}^{\infty}\frac{1}{r}b_{-r}\bar{b}_{-r}}\alpha^{\theta_0}
\left|0\right>\label{eq10}\end{equation}
where the $b$'s are the boson creation operators and $\alpha^{\theta_0}$ creates a boson in the ground state at the indicated angle, $\theta_0$, as in \cite{[11]}. The origin of the states in the twisted sector is that, the antiperiodic boundary conditions , $X\left(\sigma+\beta\right)\sim-X\left(\sigma \right)$ ($\beta$ being the period) for a closed string are allowed by the  orbifold identification. The corresponding construction for the case of Neumann boundary condition gives
\begin{equation} \left|\left|N;\theta_0\right>\right>_{tw}=\frac{1}{\sqrt{2}}e^{\sum_{r=1/2}^{\infty}\frac{1}{r}b_{-r}\bar{b}_{-r}}
\left[\alpha^0+e^{i\sqrt{2}\theta_0} \alpha^{\frac{\pi}{\sqrt{2}}}\right]\left|0\right>\label{eq11}\end{equation} with $\theta_0 = 0,\frac{\pi}{\sqrt{2}}$.

At the self dual radius, there are boundary states labeled by any elements of $SU\left(2\right)$. In those terms the orbifold identification $X\mapsto -X$ becomes $g \mapsto \sigma_{1} g \sigma_{1}$. In our parametrization it is the identification $(a,b)\mapsto (a^*,-b^*)$. A generic brane of the orbifold is a symmetric projection:
$\left|\left|g(a,b)\right>\right>_{orb}=\frac{1}{\sqrt{2}}(\left|\left|g(a,b)\right>\right>_{circ}+\left|\left|g(a^*,-b^*)\right>\right>_{circ})$
The response of this brane to radius change depends again on the bulk-boundary coupling between the bulk radius changing operator $\lambda\Phi(z,\bar{z})=\lambda J^3\bar{J}^3$ and any boundary operator $J^\gamma$ which generates infinitesimal change in the boundary conditions. Note that $\Phi$ is invariant under the orbifold identification. Analogously to (\ref{eq3}) this bulk-boundary coupling is determined by the correlator
\begin{equation}\frac{1}{2} \left(\left<J^\gamma\left(x\right)\left(J^3\bar{J}^3\right)\left(z\right)\right>_{g}+
\left<J'^\gamma\left(x\right)\left(J^3\bar{J}^3\right)\left(z\right)\right>_{g'}\right)\label{eq12}\end{equation}
where $J'^\gamma$ corresponds to the generator $t'^\gamma = \sigma^1 t^\gamma \sigma^1$ and $g'=\sigma^1 g \sigma^1$ . As in the discussion following (\ref{eq4}) the contribution of the first term is proportional to $Tr (t^\gamma [t^3,gt^3g^{-1}])$.
The contribution of the second term is proportional to $Tr(t'^\gamma[t^3,g't^3g'^{-1}])= Tr(\sigma^1 t^\gamma \sigma^1 [t^3 ,\sigma^1 g \sigma^1 t^3 \sigma^1 g^{-1} \sigma^1)$
Using the fact that $\sigma^1 t^3 \sigma^1=-t^3$ and the properties of Tr we find that the two terms are identical.
The bulk boundary coupling is proportional to $Tr (t^\gamma [t^3,gt^3g^{-1}])$ as in the case of the circle and the brane flows either to the Dirichlet condition at $b=0$ or to the Neumann condition at $a=0$.

At the fixed point set of the orbifold identification, $a$ real and $b$ imaginary, which is a circle parameterized as $a+b=e^{i\psi}$, the above brane actually represents two images which can move away from the fixed set \cite{[8]}. In addition there are twisted branes confined to the fixed set. For a general radius the twisted branes are confined to the two fixed points on the circle. The continuum of twisted branes for any $a+b=e^{i\psi}$   exists only for special radii. The above analysis shows that turning on the radius changing operator $\lambda J^{3}\bar{J}^{3}$ they either flow to $b=0,a=\pm{1}$ which are the two Dirichlet twisted states (\ref {eq10}) or to $a=0,b=\pm{1}$ which are the two Neumann twisted branes (\ref {eq11}).

A $\mathbb{Z}_{2}$ orbifold of radius $r=\frac{M}{N}r_{s.d}$ can be considered as a (non-abelian) orbifold of the self dual circle, equivalent to $k=1$ $ SU(2)$ WZW model, by the group generated by $g\mapsto e^{i\frac{\pi}{N}\sigma_{3}}ge^{-i\frac{\pi}{N}\sigma_{3}}$, $g\mapsto e^{i\frac{\pi}{M}\sigma_{3}}ge^{i\frac{\pi}{M}\sigma_{3}}$ and $g\mapsto \sigma^{1} g \sigma^{1}$. Generic boundary states are symmetric combinations of all images of the brane $\left|\left|g\right>\right>$ under this group. For various surfaces of subgroups there are also twisted branes. Since the radius changing operator $J^{3}\bar{J}^{3}$ is invariant under the full identification group, this operator will again induce on this brane the same flow as that of the case of the orbifold with self dual radius. This means that the branes which are untwisted under the $\mathbb{Z}_{2}$ operations will either flow to Dirichlet branes with $b=0$ that are combinations of a Dirichlet brane at $a$ and at $a^*$, or to Neumann branes at $a=0$ that are combinations of branes sitting at $b$ and $-b^*$. The branes that are twisted under the $\mathbb{Z}_{2}$ action will flow under radius changing to either twisted Dirichlet branes with $b=0$ and $a=\pm 1$ or to twisted Neumann branes with $a=0$ and $b=\pm i$.

 For the circle theory at the self dual radius, which is a $k=1$ $SU(2)$ WZW model (see for example \cite{[12]}), there is apparently a continuum of marginal bulk operators: every combination of $J^{1}\bar{J}^{1}$, $J^{2}\bar{J}^{2}$ and $J^{3}\bar{J}^{3}$. However each such combination is equivalent under the $SU(2)\times SU(2)$ symmetry of the model to the operator $J^{3}\bar{J}^{3}$. Turning on any such marginal combination is therefore equivalent to changing the radius by the operator $J^{3}\bar{J}^{3}$. The circle theory with a generic rational radius $r=\frac {M}{N}r_{s.d}$ is, as discussed in the previous section,  an orbifold of the self dual circle theory by the identifications $g\mapsto e^{i\frac{\pi}{N}\sigma_{3}}ge^{-i\frac{\pi}{N}\sigma_{3}}$, $g\mapsto e^{i\frac{\pi}{M}\sigma_{3}}ge^{i\frac{\pi}{M}\sigma_{3}}$. This identification breaks the $SU(2)\times SU(2)$ symmetry down to $U(1)\times U(1)$. The only marginal bulk operator consistent with this identification is the operator $J^{3}\bar{J}^{3}$ which is indeed the radius changing operator. An exception is the case $N=2, M=1$ (or its T-Dual $M=2, N=1$) for which the orbifold identification becomes $g\mapsto \sigma_{3} g \sigma_{3}$. Here on top of the radius changing operator $J^{3}\bar{J}^{3}$, the orbifold identification is also consistent with any combination of $J^{1}\bar{J}^{1}$ and $J^{2}\bar{J}^{2}$. Each such a combination is equivalent to $J^{1}\bar{J}^{1}$ under the $U(1)$ subgroup of $SU(2)$ which survives the breaking of $SU(2)$ by the orbifold identification. However, since the full $SU(2)$ is broken, the marginal operator $J^{1}\bar{J}^{1}$ is not equivalent to the radius changing operator $J^{3}\bar{J}^{3}$. So for a circle theory with $r=\frac{1}{2} r_{s.d}$ (or $r=2 r_{s.d}$) there are two inequivalent marginal bulk operators, the radius changing operator $J^{3}\bar{J}^{3}$ and another operator $J^{1}\bar{J}^{1}$. The moduli space opened by turning on this second operator is equivalent to the moduli space of the $\mathbb{Z}_{2}$ orbifold. This can be seen by mapping  the original matrix variable $g_{circ}$ to $g_{orb}=hg_{circ}h^{-1}$ with $h=\frac{1}{\sqrt{2}}(\sigma^{3} +\sigma^{1})$. This conjugation rotates $\sigma^{3}$ into $\sigma^{1}$. In terms of the variable $g_{orb}$ the orbifold identification of the circle becomes $g\mapsto \sigma_{1} g \sigma_{1}$ which is the identification of the $\mathbb{Z}_{2}$ orbifold and the operator $J^{1}\bar{J}^{1}$ becomes the radius changing operator $J^{3}\bar{J}^{3}$ for the orbifold (see figure 1).

 \setlength{\unitlength}{1mm}
 \begin{figure}[b]
  \begin{tikzpicture}
   \put(10,-105){\circle{1.6}}
   \put(9,-110){$0$}
   \put(11,-105){\line(1,0){120}}
   \put(131,-106){$\wr\wr$}
   \put(133,-105){\vector(1,0){20}}
   \put(155,-110){$r_{circ}$}
   \put(70,-105){\circle*{1.6}}
   \put(70,-105){\line(0,1){60}}
   \put(68,-46){$\sim $}
   \put(68,-45){$\sim $}
   \put(70,-44){\vector(0,1){20}}
   \put(60,-25){$r_{orb}$}
   \put(70,-105){\line(0,-1){20}}
   \put(70,-126){\circle{1.6}}
   \put(65,-126){$0$}
   \put(30,-105){\circle*{1.6}}
   \put(27,-110){$1/\sqrt{2}$}
   \put(27,-115){$SU(2)$}
   \put(55,-105){\circle*{1.6}}
   \put(54,-110){$r_1$}
   \put(45,-115){$U(1)\times U(1)$}
   \put(70,-55){\circle*{1.6}}
   \put(65,-55){$r_2$}
   \put(74,-56){$U(1)\times U(1)$}
   \put(73,-110){$\sqrt{2}$}
   \put(55,-100){$1/\sqrt{2}$}
   \put(65,-101){\vector(1,-1){4}}
   \put(75,-100){$SU(2)/ \mathbb{Z}_2$}
   \put(75,-101){\vector(-1,-1){4}}
  \end{tikzpicture}
  \caption {Moduli space of $c=1$ CFT's} The groups are the boundary-state moduli spaces at the respective radii.\\ $r_1$ and $r_2$ are generic radii which are irrational multiples of $r_{s.d}$
 \end{figure}

\pagebreak
   The point $r=r_{s.d}/2$ of the circle moduli (which is equivalent to the point $r=r_{s.d}$ of the orbifold moduli) is therefore the intersection of two different one dimensional moduli. We have seen that a generic boundary state at this point is flowing either to Dirichlet brane $(b=0 ,|a|=1)$ or to a Neumann brane $(a=0, |b|=1)$ when $J^{3}\bar{J}^{3}$ is turned on. Conjugating with the matrix $h$ we find that turning on the other operator, $J^{1}\bar{J}^{1}$ , a generic brane will flow either into $(a=a^*, b=-b^*)$ or into $(b=b^* ,a=-a^*)$. The boundary states that are stable under both moves are the eight states corresponding to $(b=0, a=\pm 1), (b=0, a=\pm i), (a=0, b=\pm 1),(a=0, b=\pm i)$.\\
  Let us add a few words about the $\mathcal{N}=1$ case. As described in \cite{[8]} the moduli space contains the following lines::
  \begin{itemize}
   \item
    The circle line.
   \item
     The orbifold line with the orbifolding operation applying to both the bosonic and fermionic coordinates.
   \item
    The super-affine line, which is derived from the circle by orbifolding a circle of radius $r$ with $S_\delta=(-1)^{F_s}e^{2\pi i p \cdot \delta}$, where $F_s$ is the fermionic number, $p=(p_L,p_R)$ and $\delta=\frac{1}{2}(r,-r)$.
   \item
    The super-orbifold which is derived by orbifolding the super-affine line.
   \item
    The orbifold-prime, derived by orbifolding the orbifold with $(-1)^{F_s}$.
  \end{itemize}
  It is shown in \cite{[13]} that the super-affine theory is equivalent to a $k=1$, $SO(3)$ WZW model. It is further shown in \cite{[8]} that one can go back to a circle of radius $r$ by acting with $S_\delta$ on a super-affine theory of radius $2r$. This, as noted in \cite{[13]} is equivalent to identifying a boundary state labeled by $g\in SO(3)$ with $\sigma^3 g \sigma^3$. As a consequence it behaves exactly like the bosonic orbifold, and therefore the same analysis holds for the circle, as well as for the orbifold lines. Since the superorbifold line is derived from the superaffine line by the orbifolding operation identifying $g$ with $\sigma^1 g \sigma^1$, then by the argument made above, it has a moduli space identical to that of the supercircle. The consequence is that in all of the above cases, a general boundary state would flow when the radius is changed to a Dirichlet or a Neumann state, just as in the bosonic case. The remaining case, that of the orbifold-prime, as stated in \cite{[8]}, carries no extra $SU(2)$ structure. Therefore, there is no induced boundary flow.

\section{Boundary-entropy analysis} \label{sec:boundary entropy}
Another consistency requirement on the result is that the boundary entropy, $s$, which is related to the ground state multiplicity, $g_b$, by $s=ln(g_b)$, decreases along all flows. In this section we show that the ground state multiplicity, $g_b$, of a general brane at a rational radius, $R_1=\frac{M}{N}R_{s.d}$ indeed decreases when the radius is changed. This generalizes the work done in \cite{[2]} from pure Dirichlet or Neumann boundary conditions to general boundary conditions labeled by elements of $SU(2)$.\\

Recall \cite{[2]}, that the ground state multiplicity of a single Dirichlet brane on a circle of radius $R$ is  $g_b=\frac{1}{\sqrt{2R}}$ while that of a Neumann brane is  $g_b=\sqrt{R}$. As shown in the previous section, a general brane in a theory compactified on a radius $R_1=\frac{M}{N}R_{s.d}$ is a result of orbifolding by the group $\mathbb{Z}_{M}\times\mathbb{Z}_{N}$ that acts on the target-space boundary conditions. Referring to the label \begin{equation}g=\left(\begin{matrix} a & b \\ -b^{*} & a^{*} \end{matrix}\right)\nonumber \end{equation} of the boundary state, the left factor of the product acts on $a$ by multiplication by an $M$'th root of unity, while the right factor acts on $b$ by multiplication by an $N$'th root of unity, as in (\ref{eq7}) and the following discussion. The brane is, therefore, a sum:
\begin{equation} \left|\left|g\left(a,b \right)\right>\right>=\frac{1}{\sqrt{MN}}
\sum\limits_{m',n'}{\left|\left|g\left(ae^{\frac{2\pi}{M} m'},be^{\frac{2\pi}{N} n'}\right)\right>\right>} \label{OrbifoldedBrane} \end{equation}
Where the sum over $m'$ is from $0$ to $M-1$, and sum over $n'$ is from $0$ to $N-1$.\\
As observed in \cite{[4]}, when the radius of the target space is increased a general brane flows to a Dirichlet brane i.e. \begin{equation} \left( \begin{matrix}  a & 0 \\ 0 &  a^{*} \end{matrix}\right)\nonumber \end{equation}. Similarly, when the radius is decreased the brane flows to the Neumann brane corresponding to the matrix \begin{equation}\left( \begin{matrix} 0 & b \\ -b^{*} &  0 \end{matrix} \right) \nonumber \end{equation}.\\
The flow of such a brane when the radius increases takes $b$ to be $0$. At this fixed point  the terms in the sum (\ref{OrbifoldedBrane}) become independent of $n'$. As a result, the sum over $n'$ in (\ref{OrbifoldedBrane}) demonstrates that one gets a stack of $N$ Dirichlet branes.\\
If we decrease the radius, the flow on (\ref{OrbifoldedBrane}) takes $a$ to be $0$. At this fixed point there is no $m'$ dependence in the terms of the sum \ref{OrbifoldedBrane}, summing over it demonstrates that one gets a stack of $M$ Neumann branes.\\

We show now how $g_b$ changes when the radius is changed. First consider the case of increasing the radius  from $R_1=\frac{M}{N} R_{s.d}$  to some $R_2 > R_1$. As mentioned above a general brane flows to a stack of $M$ Dirichlet branes. Initially, at $R_1$, the  generic  brane is described by eq. (\ref{OrbifoldedBrane}).  Since the parameters $a$ and $b$ are   moduli , $g_b$ should not depend on them \cite{[10]}. In particular  the ground state multiplicity for general  $(a,b)$ should equal to its value for ,say,  $b=0$. There eq. (\ref{OrbifoldedBrane}) describes a stack of $M$ Dirichlet branes, whose multiplicity is then  $g_b= \frac{M}{\sqrt{2R_1}} = \frac{\sqrt{MN}}{2^{1/4}}$. The same value is obtained at the point $a=0$, where (\ref{OrbifoldedBrane}) describes a stack of $N$ Neumann branes  whose ground state multiplicity is  $g_b= N \sqrt{R_1} = \frac{\sqrt{MN}}{2^{1/4}}$. This is then the initial $g_b$. The final  fixed point  of the flow resulting from increasing the radius is a stack of $M$ Dirichlet branes on a circle of radius $R_2 > R_1$. The corresponding ground state multiplicity is $g_b= \frac{M}{\sqrt{2R_2}}< \frac{M}{\sqrt{2R_1}}$. So  $g_b$ indeed decreases as required.\\
Decreasing the radius from $R_1$ to $R_2 < R_1$, we have seen that the fixed point of the corresponding flow is a stack of $N$ Neumann branes  on the circle of radius $R_2$.  The final ground state multiplicity is now  $g_b= N \sqrt{R_2} < N \sqrt{R_1}$, so the ground state multiplicity decreases again.

\section{Conclusions}
We have generalized the result that under a change of the radius general boundary states flow either to Dirichlet or Neumann (depending on whether the radius is increased or decreased), to the orbifold branch of the $c=1$ theory. We showed that this result holds away from criticality. It is also noted that the $\mathcal{N}=1$, $\hat{c}=1$ case is not very different. As a further confirmation, the final ground state multiplicity of the flow was compared to the initial one  and shown to decrease as the radius of compactification is changed.

\section*{Acknowledgments}
The writes would like to thank D. Friedan for helpful correspondence and remarks.\\
The work of S. Elitzur is partially supported by the Israel Science Foundation Center of Excellence.\\
The work of B. Karni and E. Rabinovici is partially supported by the American-Israeli Bi-National Science Foundation and the Israel Science Foundation Center of Excellence.

\end{document}